\documentclass[aps,prb,twocolumn,showpacs,showkeys,amsmath,amssymb]{revtex4}

\usepackage{epsfig,amsmath,amssymb,color}
\usepackage{graphicx}
\begin{document}

\title{Frustrated quantum Heisenberg double-tetrahedral\\
and octahedral chains at high magnetic fields}

\author{Olesia Krupnitska}

\affiliation{Institute for Condensed Matter Physics, National Academy of Sciences of Ukraine,\\ Svientsitskii Street 1, 79011 L’viv, Ukraine}

\date{\today}



\begin{abstract}
We consider the spin-1/2 antiferromagnetic Heisenberg model on two one-dimensional
frustrated lattices, double-tetrahedral chain and octahedral chain, with almost dispersionless (flat) lowest magnon
band in a strong magnetic field. Using the localized magnons picture, we construct an effective description of the one-dimensional chains with triangle and square traps within
the strong coupling approximation. We perform extensive exact diagonalization and density matrix renormalization group calculations to check the validity
of the obtained effective Hamiltonians by comparison with the initial models with special focus on the magnetization and specific heat at high magnetic fields.
\end{abstract}

\maketitle

\section{Introduction}

\label{sec1}

\setcounter{equation}{0}

The study of frustrated quantum Heisenberg antiferromagnets remains to be a hot topic in condensed matter physics \cite{FA, taki2}. Special class of frustrated quantum Heisenberg antiferromagnets, which have a dispersionless (flat) band in the one-magnon energy spectrum, is of great interest, since such systems can be mapped on the classical lattice gas of hard-core objects, see, e.g., review that has appeared recently \cite{Derzhko2015}. In  previous studies it was shown that the high-field low-temperature properties of such spin systems can be studied in detail within the localized-magnon approach \cite{locmag1, locmag2, locmag3}. The theory developed in references \cite{locmag1, locmag2, locmag3} works perfectly well in the case of so-called perfect geometry, when the one-magnon
states are strictly localized i.e., the one-magnon band is
strictly dispersionless. Moreover, the ground-state characteristic features of such spin systems are a plateau and a jump to the saturation in magnetization curve \cite{locmag1}, the spin-Peierls instability \cite{Richter2004},
and the residual entropy at the saturation field \cite{locmag2, locmag3}. However, in real systems, the conditions that provide a strict
localization of magnons can be violated and a theory of almost flat band systems is necessary. 

\begin{figure}
\begin{center}
\includegraphics[clip=on,width=80mm,angle=0]{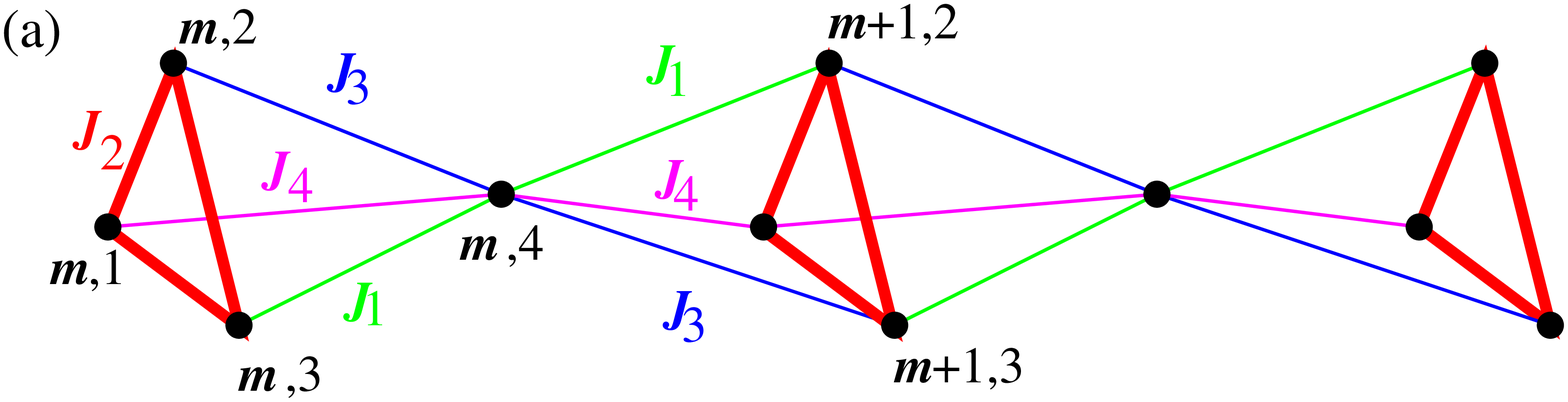}
\includegraphics[clip=on,width=80mm,angle=0]{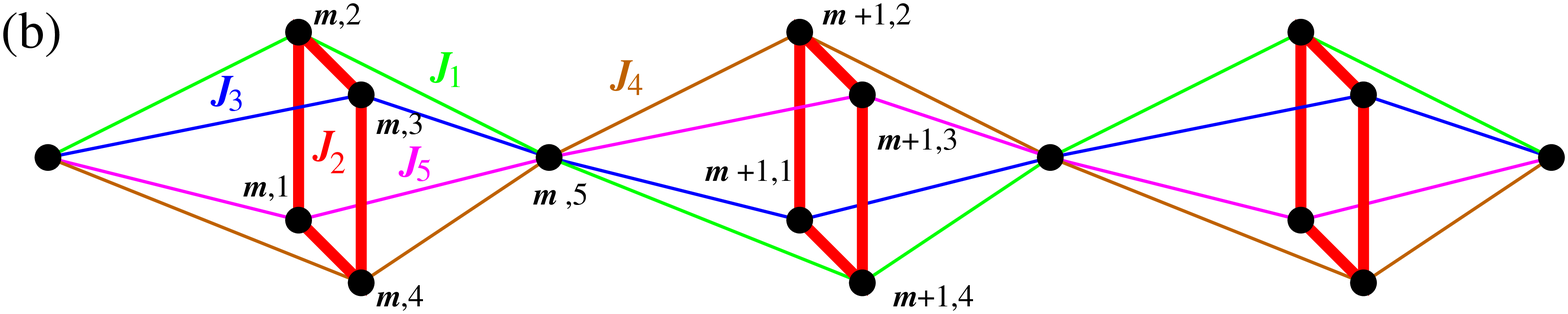}
\caption
{(Color online) The schematic of double-tetrahedral chain (a) and the octahedral chain  (b) described by Hamiltonian (\ref{001}).}
\label{fig1}
\end{center}
\end{figure}

Effects of small deviation from ideal flat-band geometry were studied in references \cite{Derzhko2013, fnt, jmmm, epl, app, tanaka}. In particular,  to study high-field low-temperature properties of initial deformed one- or two-dimensional frustrated Heisenberg antiferromagnets effective Hamiltonians were constructed using the localized-magnon approach. In the present work, we propose a systematic theory for some other systems near flat band point. To be specific, we consider two frustrated quantum spin lattices in a strong magnetic field, the double-tetrahedral chain and the octahedral chain (see Fig.~\ref{fig1}). These frustrated lattices were studied previously in the literature 
by various authors. In particular, the double-tetrahedral chain, which can be understood as a one-dimensional analog of the pyrochlore antiferromagnet, was studied in references \cite{mambrini, rojas, batista, antonosyan, chiral1, chiral2, Strecka_tetrahedral, Galisova}. Exact ground and excited states of an antiferromagnetic quantum octahedral chain were studied in papers \cite{Bose89, Bose90, Bose92} as well as low-temperature thermodynamics or quantum phase transitions of octahedral chain were investigated in works \cite{octahedral1, octahedral2}. 

It is important to note, that double-tetrahedral geometry is realized for the spin system of Cu$^{2+}$ ions in a magnetic compound Cu$_3$Mo$_2$O$_9$ \cite{Has08,Kur10,Mat12}. As shown in works \cite{Has08,Kur10,Mat12}, the exchange interaction network of this compound comprises spin-$1/2$ antiferromagnetic uniform chains $J_4$ and antiferromagnetic dimers $J_3$ which are the main determinants of the magnetism. The exchange interactions $J_1$ and $J_2$ connect the chains and dimers. A set of exchange interaction parameters $J_1$-$J_4$ estimated from the experimental data and have the following values: $J_4=6.5$ meV$\approx 75.43$ K; $J_3=5.7$ meV$\approx 66.15$ K; $J_1-J_2=3.06$ meV$\approx 35.51$ K \cite{Mat12}. { It is worth noting a whole class of geometrically frustrated compounds based on cobalt oxide, RBaCo$_4$O$_7$ with a rare earth atom R, which has swedenborgite lattice structure \cite{Buhrandt}. Considering one columnar stripe in such magnetic compound we could single out a double-tetrahedral chain structure.} On the other hand, the solid-state realization of the octahedral spin chain have not been found yet, { but spin clusters with a geometric shape of octahedron can be found in the polynuclear complexes such as Cu$_6$ \cite{Liu} or V$_6$ \cite{Daniel}. We believe that our theoretical findings could be useful in synthesis of solid realization of the octahedral spin chain by experimentalists.}

In what follows, we consider the spin-1/2 antiferromagnetic Heisenberg model on double-tetrahedral chain  and the octahedral chain (see Fig. \ref{fig1}) in a magnetic field
with the Hamiltonian
\begin{eqnarray}
\label{001}
H=\sum_{(ij)} J_{ij} {\bf{s}}_i \cdot {\bf{s}}_j-hS^z,
\;\;\;
S^z=\sum_{i=1}^Ns_i^z.
\end{eqnarray}
Here the first sum runs over all nearest neighbors on a lattice, whereas the second one runs over all $N$ lattice sites.
Note that $[S^z,H]=0$, i.e., the eigenvalues of $S^z$ are good quantum numbers. To accomplish description of the model,
we introduce a convenient labeling of the lattice sites by a pair of indices,
where the first number enumerates the cells
($m=1,\ldots,{\cal{N}}$, ${\cal{N}} = N/4$ for the double-tetrahedral chain or ${\cal{N}} = N/5$ for the octahedral chain, $N$ is the number of sites)
and the second one enumerates the position of the site within the cell, see Fig. \ref{fig1}.
Both these models may support independent localized-magnon states, which dominate their high-field low-temperature thermodynamics.
A localized magnon state can be located within a trapping cell (equilateral triangle or square) due to destructive quantum interference.
It has the lowest energy in the subspace $S^z=N/2-1$ if the strength of the antiferromagnetic bonds of the trapping cells $J_2$ exceeds a lower bound \cite{locmag1}. Owing to the localized nature of these states the many-magnon ground state can be constructed by filling the traps by localized magnons. Moreover, magnon localization occurs due to the specific lattice geometry
and hence requires a certain relations between the exchange interactions $J_{ij}$.
For the considered traps (triangle or square) this condition is fulfilled
if an arbitrary bond of the trapping cell and the surrounding bonds attached to the two sites of this bond form an isosceles triangle,
i.e. $J_{i}=J, i\neq2$ in Fig. \ref{fig1}. 
In the present study we deal with the case when the localization conditions are slightly violated.

In previous investigations it was shown, that localized states on trapping cells with even number of sites (vertical bond for diamond or dimer-plaquette chains, square for square-kagome lattice) have a
nondegenerate ground state for the one-particle problem \cite{Derzhko2013, fnt, jmmm}. A new interesting feature concerns to the frustrated double-tetrahedral chain. In general, lattices with odd sites in trapping cells has an additional degree of freedom -- chirality \cite{chirality}, which leads to increasing of the degeneracy of the ground state of the double-tetrahedral chain. That is why one of the main goals of present research is the study of interplay between chirality and non ideal geometry of the lattice and their influence on experimentally observed thermodynamic quantities.

The paper is organized as follows. In Sec.~\ref{sec2} we construct an effective Hamiltonians for the double-tetrahedral chain and the octahedral chain in a strong magnetic field within strong-coupling approximation. In Sec.~\ref{sec3} we compare exact diagonalization and density matrix renormalization group results for the initial and the corresponding effective models to estimate the validity of the obtained effective models and discuss the magnetization process and temperature dependence of specific heat. We summarize our findings in Sec.~\ref{sec4}.

\section{Effective Hamiltonians. Strong-coupling approach}

\label{sec2}

\setcounter{equation}{0}

To study the low-energy behavior of the introduced models we use a strong-coupling perturbation theory. For considered systems
one can easily single out a regular pattern of ${\cal{N}}$ trapping cells which do not have common sites and have sufficiently large couplings $J_2$
(the squares for the octahedral chain or the equilateral triangles for the double-tetrahedral chain). The traps are joined via weaker connecting bonds $J_i$, $i\ne 2$,
which in the ideal geometry case prevent the escape of the localized magnons from the traps.

Within the strong-coupling approach one assumes that the coupling $J_2$ is the dominant one,
i.e., $J_i/J_2\ll 1$, $i\ne 2$.
At high fields only few states of the trapping cell are relevant,
namely,
the fully polarized state $\vert u\rangle$ and the localized-magnon state $\vert d\rangle$; in the case of the triangular trap there are two localized-magnon states $\vert d+\rangle$ and $\vert d-\rangle$ with different chiralities. All other sites $m,4$ or $m,5$, $m=1,...,{\cal{N}}$ have fully polarized spins.
As the magnetic field decreases from very large values,
the ground state of the trap undergoes a transition between the state $\vert u\rangle$ and the state $\vert d\rangle$ at the ``bare'' saturation field $h_0$.
The Hamiltonian $H$ is splitted into a ``main'' part $H_{\rm{main}}$
(the Hamiltonian of all traps and the Zeeman interaction of all spins with the magnetic field $h_0$)
and a perturbation $V=H-H_{\rm{main}}$.
The ground state $\vert\varphi_0\rangle$ of the Hamiltonian without the connecting bonds $J_i=0$, $i\ne 2$ and at $h-h_0=0$
is $2^{{\cal{N}}}$-fold degenerate for the case of octahedral chain and $3^{{\cal{N}}}$-fold degenerate in the case of the double-tetrahedral chain. These ground states form a model subspace defined by the projector $P=\vert\varphi_0\rangle\langle\varphi_0\vert$.
When $J_i$, $i\ne 2$ and $h-h_0$ deviate from zero
we are interested in an effective Hamiltonian $H_{\rm{eff}}$ which acts on the model space only but gives the exact ground-state energy.
$H_{\rm{eff}}$ can be found perturbatively \cite{klein,fulde,essler} and is given by
\begin{eqnarray}
\label{002}
H_{\rm{eff}}
=PHP+PV\sum_{\alpha\ne 0}\frac{\vert \varphi_{\alpha}\rangle\langle \varphi_{\alpha}\vert}{\varepsilon_0-\varepsilon_{\alpha}}VP+\ldots.
\end{eqnarray}
Here $\vert \varphi_{\alpha}\rangle$, $\alpha\ne 0$ are excited states of $H_{\rm{main}}$.
To rewrite the effective Hamiltonian in a more transparent form amenable for further analysis it might be convenient to introduce (pseudo)spin operators representing the states of each trapping cell.

\subsubsection{Heisenberg double-tetrahedral chain}

A new feature in the case of the double-tetrahedral chain is related to the chirality \cite{chirality} of the triangle.
As a result, in a strong magnetic field for each triangle we consider three states:
\begin{eqnarray}
\label{003}
\begin{split}
&\vert u\rangle=\vert \uparrow\uparrow\uparrow\rangle,
{}\\&
\vert d+\rangle=\frac{1}{\sqrt{3}}\left(\vert \downarrow\uparrow\uparrow\rangle +\omega\vert \uparrow\downarrow\uparrow\rangle +\omega^2\vert \uparrow\uparrow\downarrow\rangle\right),
{}\\&
\vert d-\rangle=\frac{1}{\sqrt{3}}\left(\vert \downarrow\uparrow\uparrow\rangle +\omega^2\vert \uparrow\downarrow\uparrow\rangle +\omega\vert \uparrow\uparrow\downarrow\rangle\right),
{}\\&
\omega=e^{i\frac{2\pi}{3}}.
\end{split}
\end{eqnarray}
Their energies are $3J_2/4-3h/2$,  $-3J_2/4-h/2$, and $-3J_2/4-h/2$, respectively.
The saturation field is $h_0=3J_2/2$. If $h=h_0$ and $J_{1}=J_{3}=J_{4}=0$ the ground state of $H$, $\vert\varphi_0\rangle$, is $3^{{\cal{N}}}$-fold degenerate.
The projector onto the ground states of $H_{\rm{main}}$ is
\begin{eqnarray}
\label{004}
\begin{split}
&P=\vert \varphi_0\rangle\langle \varphi_0\vert=\otimes_{m}P_{m},
{}\\&
P_m=\left(\left(\vert u\rangle\langle u\vert+\vert d+\rangle\langle d+\vert + \vert d-\rangle\langle d-\vert\right)
\otimes \vert \uparrow_4\rangle\langle \uparrow_4\vert\right)_m.
\end{split}
\end{eqnarray}

Evidently, we face a spin-$1$ problem.
We use the following representation for the (pseudo)spin-1 operators:
\begin{eqnarray}
\label{005}
\begin{split}
&T^x
=
\frac{1}{\sqrt{2}}\left(
\begin{array}{ccc}
0 & 1 & 0 \\
1 & 0 & 1 \\
0 & 1 & 0 \\
\end{array}
\right),
\,\,\,
T^y
=
\frac{1}{i\sqrt{2}}\left(
\begin{array}{ccc}
0 & 1 & 0 \\
-1 & 0 & 1 \\
0 & -1 & 0 \\
\end{array}
\right),
{}\\&
T^z
=
\left(
\begin{array}{ccc}
1 & 0 & 0 \\
0 & 0 & 0 \\
0 & 0 & -1 \\
\end{array}
\right).
\end{split}
\end{eqnarray}
Let us put
\begin{eqnarray}
\begin{split}
&\vert d+\rangle\langle d+\vert\to
\left(
\begin{array}{c}
1  \\
0  \\
0  \\
\end{array}
\right)
\left(
\begin{array}{ccc}
1 & 0 & 0
\end{array}
\right)
{}\\&
=
\left(
\begin{array}{ccc}
1 & 0 & 0 \\
0 & 0 & 0 \\
0 & 0 & 0 \\
\end{array}
\right)=\frac{1}{2}\left(T^z+{T^z}^2\right),
{}\\&
\vert u\rangle\langle u\vert\to
\left(
\begin{array}{ccc}
0 & 0 & 0 \\
0 & 1 & 0 \\
0 & 0 & 0 \\
\end{array}
\right)=1-{T^z}^2,
{}\\&
\vert d-\rangle\langle d-\vert\to
\left(
\begin{array}{ccc}
0 & 0 & 0 \\
0 & 0 & 0 \\
0 & 0 & 1 \\
\end{array}
\right)=\frac{1}{2}\left(-T^z+{T^z}^2\right).
\end{split}
\end{eqnarray}
We have to notice here that such an encoding contradicts the intuitive expectation  ``the fully polarized state corresponds to $T^z=+1$''.
According to the introduced correspondence, the fully polarized state corresponds to $T^z=0$.
Indeed, while $s$ is real spin which interacts with the magnetic field, ${\bf T}$ is a pseudospin which simply represents three states of the triangle.

Then the first term in Eq. (\ref{002}) can be written in the terms of (pseudo)spin-1 operators:
\begin{eqnarray}
\label{006}
\begin{split}
&PHP=P\sum_{m=1}^{{\cal{N}}}
\left(-2h+\frac{3}{4}J_2+\frac{3J}{2}
+\left(h-h_1 \right){T_m^z}^2
\right)P,
{}\\&
h_1=\frac{3}{2}J_2 +J, 
\,\,\,
J=\frac{J_{1}+J_{3}+J_{4}}{3}.
\end{split}
\end{eqnarray}
For the ideal geometry case this is the well known result \cite{chiral2}.

Relevant excited states contain only one flipped spin on the site connecting two neighboring triangles.
The energy of these excited states is $\varepsilon_0+h_0$ with $h_0=3J_2/2$ and therefore for the second term in Eq. (\ref{002}) we have:
\begin{eqnarray}
\begin{split}
\label{007}
&PV\sum_{\alpha\ne 0}\frac{\vert\varphi_{\alpha}\rangle\langle \varphi_{\alpha}\vert}{\varepsilon_0-\varepsilon_{\alpha}}VP
{}\\&
=
-\frac{2}{3J_2}
\sum_{m=1}^{{\cal{N}}}
P\left(
\frac{J_{4}}{2} s_{m,1}^- +\frac{J_{3}}{2} s_{m,2}^- +\frac{J_{1}}{2} s_{m,3}^-
\right.
{}\\&
\left.
+ \frac{J_{4}}{2}s_{m+1,1}^- + \frac{J_{1}}{2} s_{m+1,2}^- + \frac{J_{3}}{2} s_{m+1,3}^-
\right)
{}\\&
\times
\left(
\frac{J_{4}}{2} s_{m,1}^+   +\frac{J_{3}}{2} s_{m,2}^+ +\frac{J_{1}}{2} s_{m,3}^+
 \right.
{}\\&
\left.
 + \frac{J_{4}}{2} s_{m+1,1}^+ +\frac{J_{1}}{2} s_{m+1,2}^+  + \frac{J_{3}}{2} s_{m+1,3}^+
\right)P.
\end{split}
\end{eqnarray}
After the evaluation of all necessary matrix elements $Ps^{\mp}_{m,i}P$, we get
\begin{eqnarray}
\label{008}
\begin{split}
&PV\sum_{\alpha\ne 0}\frac{\vert\varphi_{\alpha}\rangle\langle \varphi_{\alpha}\vert}{\varepsilon_0-\varepsilon_{\alpha}}VP
{}\\&
=-\frac{1}{18J_2}\sum_{m=1}^{{\cal{N}}}
P\left( \left(a\vert d+\rangle +a^*\vert d-\rangle\right)\langle u\vert_m
\right.
{}\\&
\left.
+ \left(a^*\vert d+\rangle +a\vert d-\rangle\right)\langle u\vert_{m+1} \right)
\times
\left( \vert u\rangle \left(a^* \langle d+\vert  + a\langle d-\vert\right)_m
\right.
{}\\&
\left.
+ \vert u\rangle \left(a \langle d+\vert  + a^*\langle d-\vert\right)_{m+1} \right)P.
\end{split}
\end{eqnarray}
Here
\begin{eqnarray}
\label{009}
\begin{split}
&a=J_{4}+\omega^2 J_{3}+\omega J_{1}=\vert a\vert e^{i\alpha},
{}\\&
\vert a\vert=\sqrt{J_{1}^2+J_{3}^2+J_{4}^2-J_{1}J_{4}-J_{1}J_{4}-J_{3}J_{4}}.
\end{split}
\end{eqnarray}\\

Using (pseudo)spin-$1$ $T^\alpha_m$ operators, we have the following expression for effective Hamiltonian:
\begin{widetext}
\begin{eqnarray}
\label{010}\begin{split}
&H_{\rm{eff}}=\sum_{m=1}^{{\cal{N}}}
\left[C+\left(h-h_1-\frac{\vert a\vert^2}{9J_2}\right){T_m^z}^2
-\frac{1}{18J_2}\left[(a^2+a^{*2})\left({T^{x}_m}^2-{T^{y}_m}^2\right)
\right.
\right.
{}\\&
\left.
\left.
+\frac{1}{2}\left(\left(a-a^*\right)^2\left(T^{z}_mT^{x}_mT^{x}_{m+1}T^{z}_{m+1}+T^{x}_mT^{z}_mT^{z}_{m+1}T^{x}_{m+1}\right)
+\left(a+a^*\right)^2\left(T^{z}_mT^{y}_mT^{y}_{m+1}T^{z}_{m+1}+T^{y}_mT^{z}_mT^{z}_{m+1}T^{y}_{m+1}\right)
\right.
\right.
\right.
{}\\&
\left.
\left.
\left.
+i\left({a^*}^2-a^2\right)\left(T^{z}_mT^{x}_mT^{y}_{m+1}T^{z}_{m+1}+T^{x}_mT^{z}_mT^{z}_{m+1}T^{y}_{m+1}\right)
+i\left(a^2-{a^*}^2\right)\left(T^{z}_mT^{y}_mT^{x}_{m+1}T^{z}_{m+1}+T^{y}_mT^{z}_mT^{z}_{m+1}T^{x}_{m+1}\right)
\right)\right]
\right],
{}\\&
C=-2h+\frac{3}{4}J_2+\frac{3J}{2},
\,\,\,
h_1=\frac{3}{2}J_2 +J, 
\,\,\,
J=\frac{J_{1}+J_{3}+J_{4}}{3}, \,\,\,
a=J_{4}+\omega^2 J_{3}+\omega J_{1},  \omega=e^{i\frac{2\pi}{3}}.
\end{split}
\end{eqnarray}
This result can be also written in more compact matrix notations: 
\begin{eqnarray}
\label{011}
\begin{split}
H_{\rm eff}
=\sum_{m=1}^{{\cal{N}}}&\left[C\left(\begin{array}{ccc}
1 & 0 & 0 \\
0 & 1 & 0 \\
0 & 0 & 1
\end{array}\right)+\left(\begin{array}{ccc}
h-h_1 & 0 & 0 \\
0 & 0 & 0 \\
0 & 0 & h-h_1 
\end{array}\right)_{m}
-\frac{1}{18J_2}
\left(
\left(\begin{array}{ccc}
2\vert a\vert^2 & 0 & a^2+{a^*}^2 \\
0 & 0 & 0 \\
{a}^2+{a^*}^2 & 0 & 2\vert a\vert^2 
\end{array}\right)_{m}
\right.
\right.
{}\\&
\left.
\left.
+\frac{1}{2}
\left(\begin{array}{ccc}
0 & a & 0 \\
0 & 0 & 0 \\
0 & a^* & 0
\end{array}\right)_{m}
\left(\begin{array}{ccc}
0 & 0 & 0 \\
a & 0 & a^* \\
0 & 0 & 0
\end{array}\right)_{m+1}
+
\frac{1}{2}
\left(\begin{array}{ccc}
0 & 0 & 0 \\
a^* & 0 & a \\
0 & 0 & 0
\end{array}\right)_{m}
\left(\begin{array}{ccc}
0 & a^* & 0 \\
0 & 0 & 0 \\
0 & a & 0
\end{array}\right)_{m+1}
\right)\right],
{}\\&
C=-2h+\frac{3}{4}J_2+\frac{3J}{2},
\,\,\,
h_1=\frac{3}{2}J_2 +J, 
\,\,\,
J=\frac{J_{1}+J_{3}+J_{4}}{3}, \,\,\,
a=J_{4}+\omega^2 J_{3}+\omega J_{1},\,\,\, \omega=e^{i\frac{2\pi}{3}}.
\end{split}
\end{eqnarray}
The obtained effective Hamiltonian acts in the space of $3^{{\cal{N}}}$ ground states of $H_{\rm{main}}$  and corresponds to unfrustrated (pseudo)spin-1 chain in magnetic field. For the ideal geometry case, i.e., when $J_1=J_3=J_4$ ($J_{4}+\omega^2 J_{3}+\omega J_{1}=0$) we have noninteracting (pseudo)spins 1 in a magnetic field.

\end{widetext}

\subsubsection{Heisenberg octahedral chain}

In the case of the octahedral chain the following two states of each square are relevant in a strong magnetic field
\begin{eqnarray}
\label{012}
\begin{split}
&\vert u\rangle= \vert\uparrow\uparrow\uparrow\uparrow\rangle,
{}\\&
\vert d\rangle= \frac{1}{2}
\left(\vert\uparrow\uparrow\uparrow\downarrow\rangle - \vert\uparrow\uparrow\downarrow\uparrow\rangle
+\vert\uparrow\downarrow\uparrow\uparrow\rangle - \vert\downarrow\uparrow\uparrow\uparrow\rangle\right).
\end{split}
\end{eqnarray}
Their energies are $J_2-2h$ and $-J_2-h$, respectively.
For the projector onto the ground-state manifold of $H_{\rm{main}}$ we have
\begin{eqnarray}
\label{013}
\begin{split}
&P=\vert \varphi_0\rangle\langle \varphi_0\vert=\otimes_{m}P_{m},
{}\\&
P_{m}=\left((\vert u\rangle\langle u\vert+\vert d\rangle\langle d\vert)
\otimes
\vert\uparrow_5\rangle\langle\uparrow_5\vert
\right)_{m}.
\end{split}
\end{eqnarray}
Similar to the double-tetrahedral chain, the set of relevant excited states $\vert\varphi_{\alpha}\rangle$, $\alpha\ne 0$,
is  the set of the states with only one flipped spin $\vert\downarrow_5\rangle$ on those sites which connect two neighboring squares $J_2$. Using (pseudo)spin-$1/2$ operators for each cell
\begin{eqnarray}
\label{014}
\begin{split}
&1=\vert u\rangle\langle u\vert + \vert d\rangle\langle d\vert,
\,\,\,
T^z=\frac{1}{2}\left(\vert u\rangle\langle u\vert - \vert d\rangle\langle d\vert\right),
{}\\&
\vert u\rangle\langle u\vert=\frac{1}{2}+T^z,
\,\,\,
\vert d\rangle\langle d\vert=\frac{1}{2}-T^z,
\end{split}
\end{eqnarray}
Eq. (\ref{002}) becomes
\begin{eqnarray}
\begin{split}
\label{015}
&H_{\rm{eff}}
=
\sum_{m}\left(C-{\sf{h}}T_{m}^z
-{\sf{J}}\left(T_{m}^xT_{m+1}^x+T_{m}^yT_{m+1}^y\right)\right),
{}\\&
C=-2h+J_2+\frac{3}{2}J-\frac{\left(J_1-J_3+J_4-J_5\right)^2}{16J_2},
{}\\&
{\sf{h}}=h-h_1-\frac{\left(J_1-J_3+J_4-J_5\right)^2}{8J_2},
{}\\&
{\sf{J}}=\frac{\left(J_1-J_3+J_4-J_5\right)^2}{16J_2}.
\end{split}
\end{eqnarray}
The obtained an effective Hamiltonian in strong-coupling approach acts in space of two ground states of $H_{\rm{main}}$ and corresponds to unfrustrated spin-$1/2$ isotropic $XY$ chain in a transverse magnetic field \cite{lieb}. Therefore the free energy (per cell) of the initial frustrated quantum spin model in the low-temperature strong-field regime for ${\cal{N}}\to\infty$ is given by
\begin{eqnarray}
\label{016}
f(T,h)={\sf{C}}-\frac{T}{2\pi}\int_{-\pi}^{\pi}{\rm{d}}\kappa \ln\left(2\cosh\frac{\Lambda_\kappa}{2T}\right)
\end{eqnarray}
with 
\begin{eqnarray}
\label{017}
\Lambda_\kappa={\sf{h}} + {\sf{J}}\cos(\kappa).
\end{eqnarray}
Knowing the free energy (\ref{016}) one can easily obtain all thermodynamic quantities. For example, the magnetization per cell is given by $M(T, h) = -\partial f(T,h)/\partial h$, as well as specific heat $C(T,h)$ can be found as $C(T,h) = -T \partial^2 f(T,h)/\partial T^2 $ (the magnetization per site and specific heat
are five times smaller: $m(T,h)=M(T,h)/5$, $c(T,h)=C(T,h)/5)$.
In the case of the ideal geometry, i.e., when $J_1=J_3=J_4=J_5$, we face with noninteracting (pseudo)spins $1/2$ in a magnetic field.

\section{High-field low-temperature thermodynamics}

\label{sec3}

\setcounter{equation}{0}

In this section we will verify the region of validity of the obtained effective Hamiltonians (\ref{011}) and (\ref{015}). To check the quality of the obtained effective description we perform exact diagonalization (ED) and density matrix renormalization group (DMRG) calculations \cite{alps}. We consider initial chains with periodic boundary conditions ($N=4{\cal{N}}=16$ for the double-tetrahedral chain and $N=5{\cal{N}}=15$ for the octahedral chain) in ED calculations or open chains with ${\cal{N}}=60$ in DMRG calculations and compute 
the magnetization $m(T,h)$ at low temperatures 
and the specific heat $c(T,h)$ at high fields. Periodic boundary conditions are applicable to systems with a small number of sites to ensure the equivalence of all sites. This reduces the volume of computations by reducing the Hilbert dimension of the entire system. We have used the DMRG method to expand our chain length beyond the exact limit. The most important difference to other numerical methods is that DMRG prefers open boundary conditions { and in present study it is used for the zero temperature}. In the evaluation of magnetization curves by DMRG method we used such control parameters: sweeps=4 and maxstates=100. Results for the initial full model (\ref{001}) are compared with corresponding data for effective models. 

\begin{figure}[h!]
\begin{center}
\includegraphics[clip=on,width=85mm,angle=0]{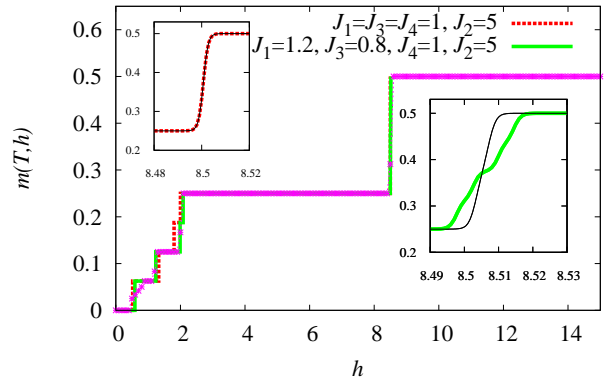}\\
\caption{(Color online) Field dependences of the magnetization (per site) $m(T,h)$ of the double-tetrahedral chain
for ideal ($J_1=J_3=J_4=1$, $J_2=5$, red dotted line, left inset) and deformed geometry ($J_1=1.2$, $J_2=5$, $J_3=0.8$, $J_4=1$, green thick line, right inset) of the lattice. ED data for the initial model $(N = 16)$ are compared with corresponding one for the effective model (${\cal N} = 4$, thin black lines). DMRG data obtained for ${\cal N} = 60$, i.e., $N=240$ (magenta symbols).}
\label{fig2}
\end{center}
\end{figure}
\begin{figure}[h!]
\begin{center}
\includegraphics[clip=on,width=85mm,angle=0]{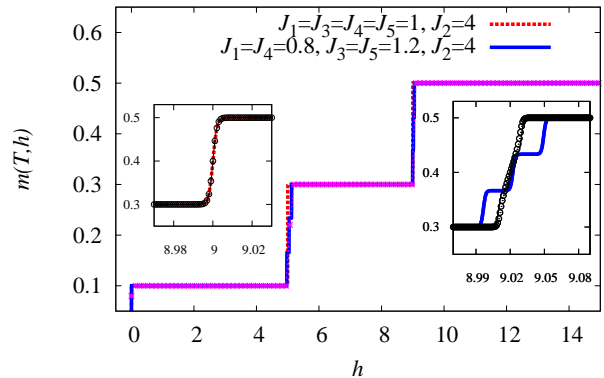}
\caption{(Color online) Field dependences of the magnetization (per site) $m(T,h)$ of the octahedral chain for ideal ($J_1=J_3=J_4=1=J_5$, $J_2=4$, red dotted line, left inset) and deformed geometry ($J_1=J_4=0.8$, $J_3=J_5=1.2$, $J_2=4$, blue thick line, right inset) of the lattice. ED data for the initial model $(N = 15)$ are compared with corresponding one for the effective model (${\cal N} = 3$, thin black lines). DMRG data obtained for ${\cal N} = 60$, i.e., $N=300$ (magenta symbols). Effective-model predictions  for thermodynamically large chains are denoted by empty black circles.}
\label{fig3}
\end{center}
\end{figure}

In { main panels of Fig.~\ref{fig2} and { Fig.~\ref{fig3}} we show the total magnetization curves at zero temperature} for the considered models of $N=16$ (the double-tetrahedral chain,  Fig.~\ref{fig2}), $N=15$ (the octahedral chain,  Fig.~\ref{fig3}). In these figures  dotted red curves indicate the ED results for ideal geometry as well as thick solid green and blue curves denote the ED results for slightly distorted geometry; magenta symbols indicate the DMRG results for initial deformed models; thin black curves indicate the results for effective models. The magnetization curves demonstrate the existence of plateaus, which are a characteristic feature of localized magnons, at one-third  of the saturation magnetization in the case of the double-tetrahedral chain (Fig.~\ref{fig2}) and at one-fifth and three-fifths of the
saturation magnetization for the octahedral chain (Fig.~\ref{fig3}). Insets demonstrate {  the influence of finite temperature, i. e. $T =0.001$ on magnetization and also show }the region of applicability of the constructed effective theory for ideal and distorted geometry {  of the lattice}. To be specific, left insets in Fig.~\ref{fig2} and Fig.~\ref{fig3} demonstrate the strict jump of magnetization to the saturation value for the ideal double-tetrahedral and octahedral chains. Comparing the results of ED calculations for initial (dotted red curves) and effective models (thin black curves) for ideal chains indicates their excellent agreement. Right insets in Fig.~\ref{fig2} and Fig.~\ref{fig3} show how these the strict jumps modify due to small deviations from the ideal geometry.
\begin{figure}[h!]
\begin{center}
\includegraphics[clip=on,width=80mm,angle=0]{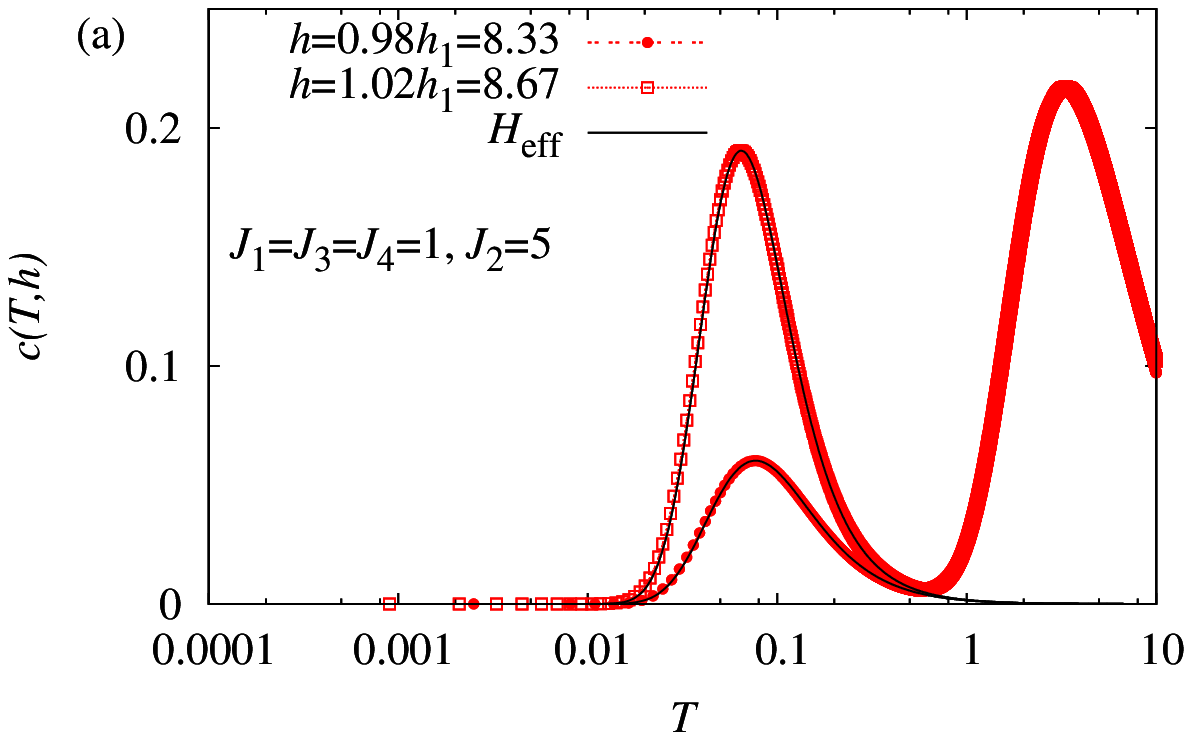}\\
\includegraphics[clip=on,width=80mm,angle=0]{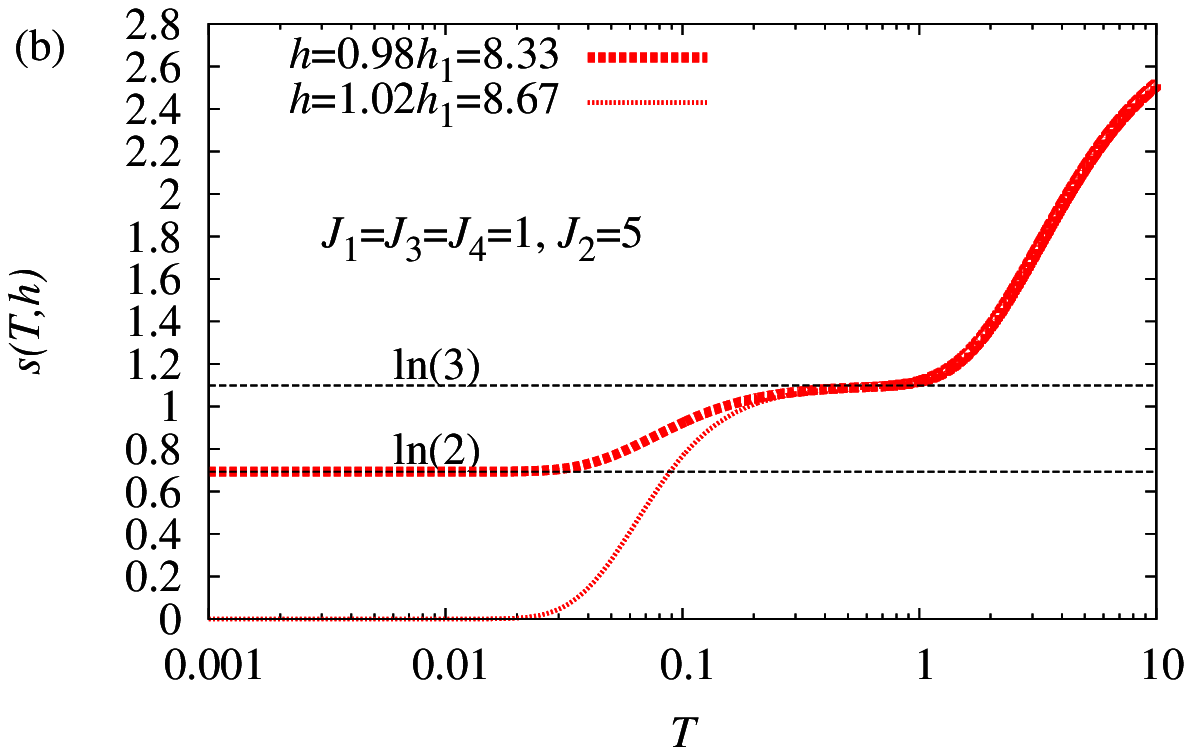}
\caption{(Color online) Temperature dependences of the specific heat (per site) $c(T,h)$ { (a) and entropy (per cell) $s(T,h)$ (b) of the double-tetrahedral chain
for ideal ($J_1=J_3=J_4=1$, $J_2=5$) geometry of the lattice.} ED data for the initial model $(N = 16)$ are compared with corresponding one for the effective model (${\cal N} = 4$, thin black curves).}
\label{fig4}
\end{center}
\end{figure}
\begin{figure}[h!]
\begin{center}
\includegraphics[clip=on,width=80mm,angle=0]{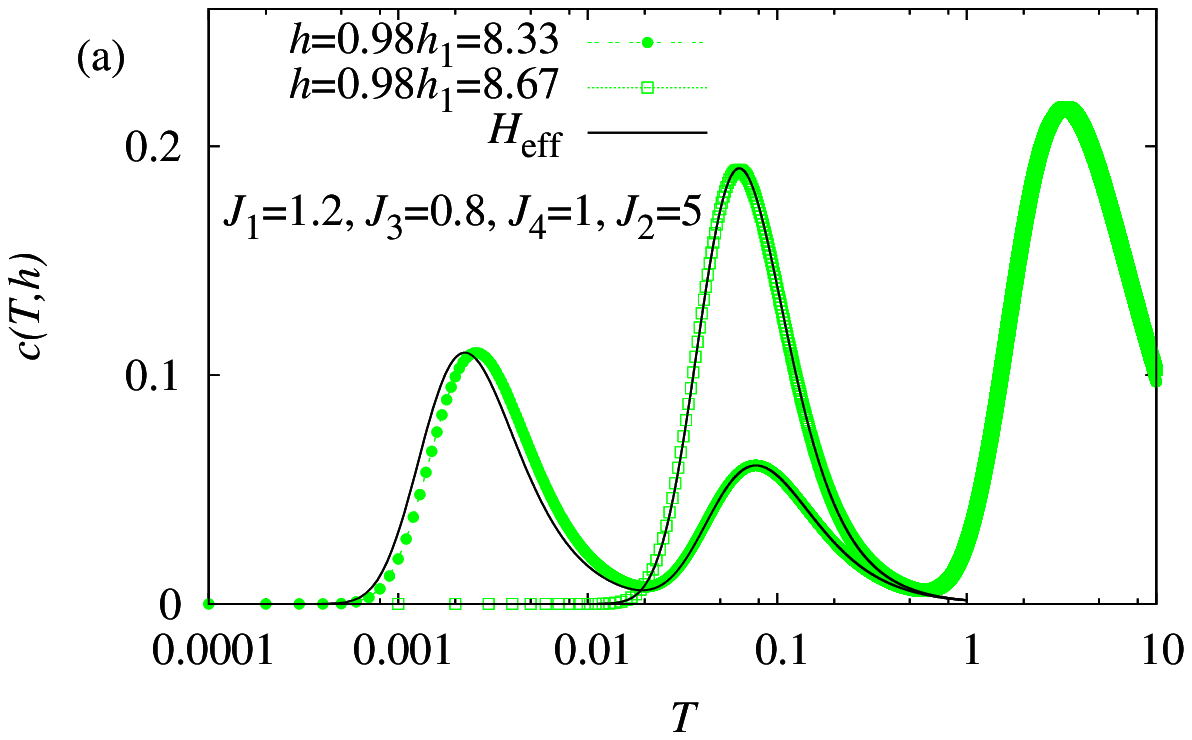}\\
\includegraphics[clip=on,width=80mm,angle=0]{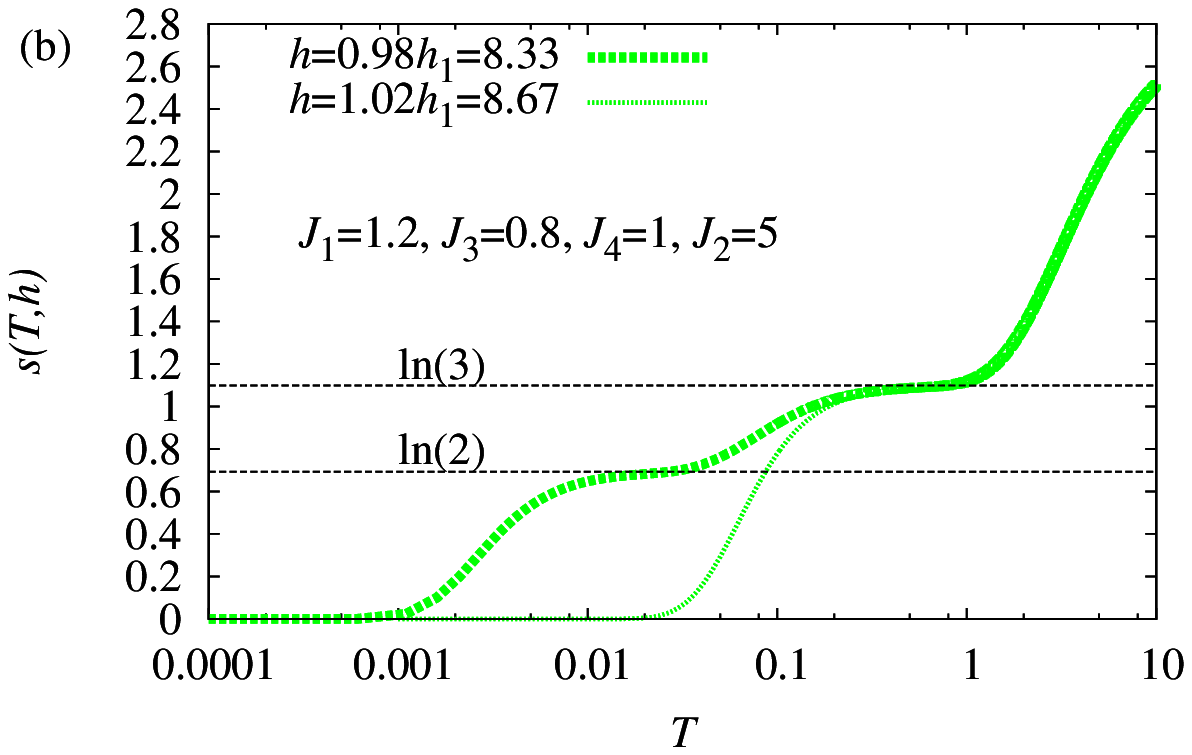}
\caption{(Color online) Temperature dependences of the specific heat (per site) $c(T,h)$ { (a) and entropy (per cell) $s(T,h)$ (b) of the double-tetrahedral chain for deformed geometry ($J_1=1.2$, $J_2=5$, $J_3=0.8$, $J_4=1$) of the lattice. } ED data for the initial model $(N = 16)$ are compared with corresponding one for the effective model (${\cal N} = 4$, thin black curves).}
\label{fig5}
\end{center}
\end{figure}

\begin{figure}[h!]
\begin{center}
\includegraphics[clip=on,width=80mm,angle=0]{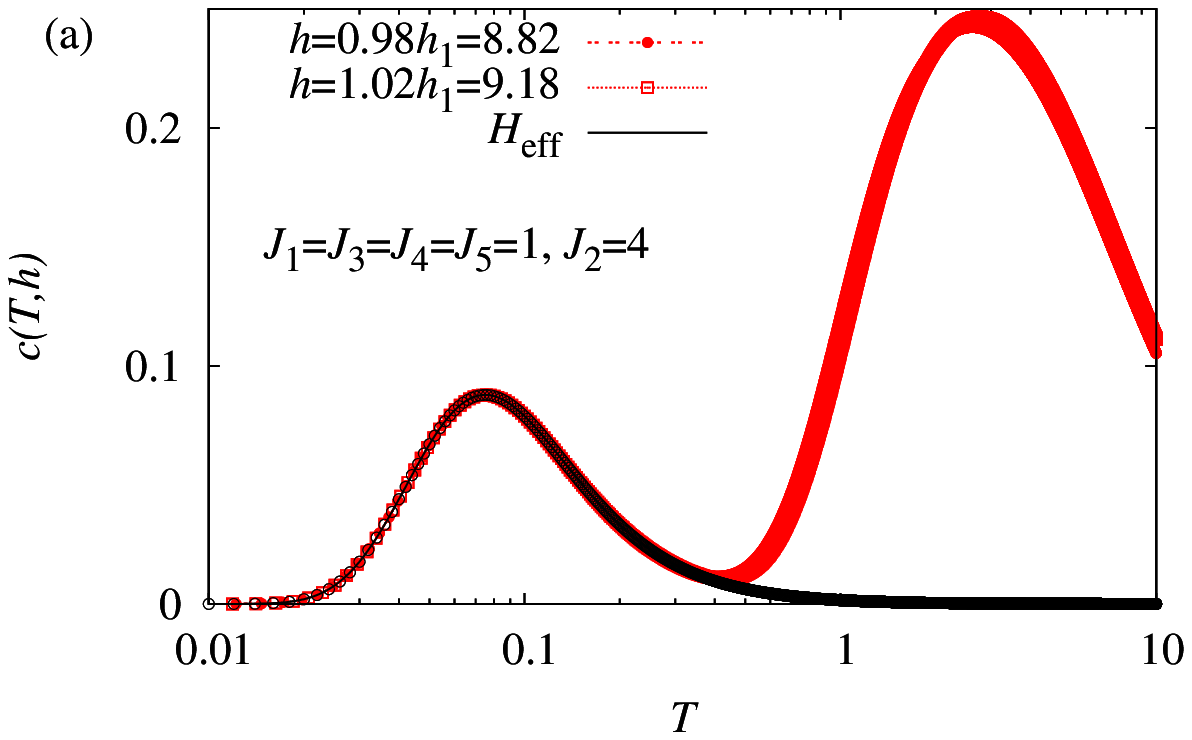}\\
\includegraphics[clip=on,width=80mm,angle=0]{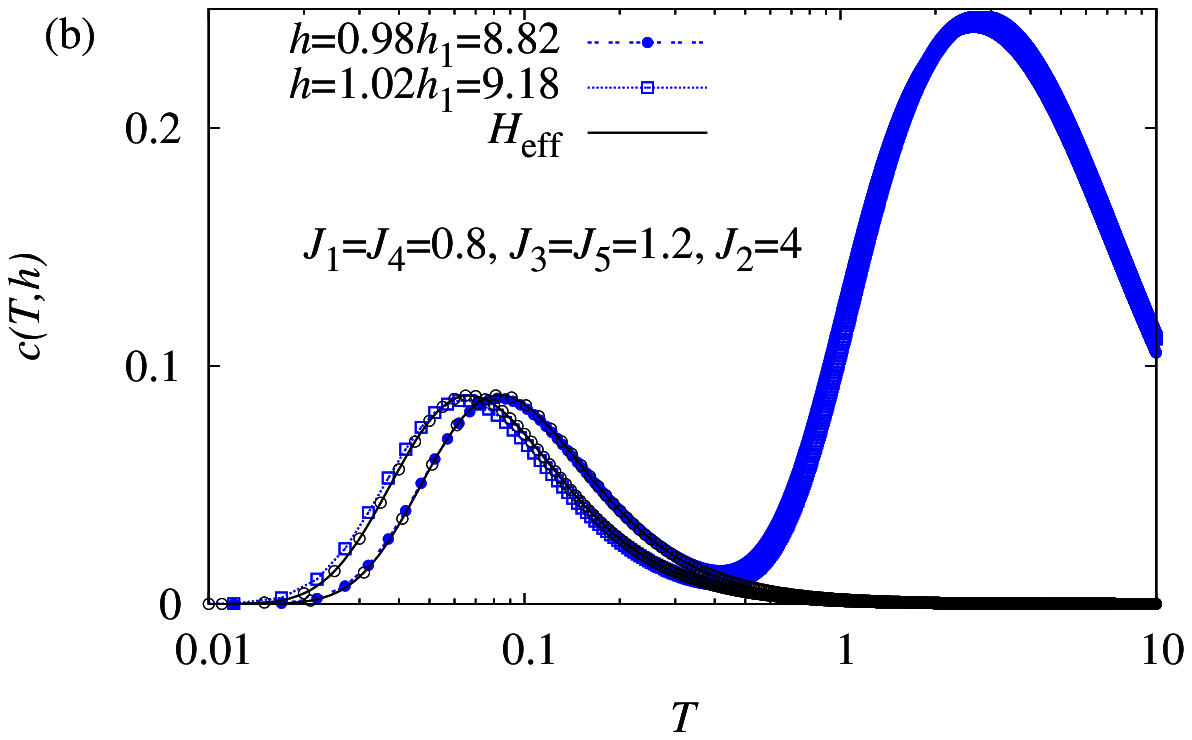}
\caption{(Color online) Temperature dependences of the specific heat (per site) $c(T,h)$ of the octahedral chain
for ideal ($J_1=J_3=J_4=J_5=1$, $J_2=4$, red symbols) (a) and deformed geometry ($J_1=J_4=0.8$, $J_3=J_5=1.2$, $J_2=4$, blue symbols) (b) of the lattice. ED data for the initial model $(N = 15)$ are compared with corresponding one for the effective model (${\cal N} = 3$, thin black lines).  Effective-model predictions  for thermodynamically large chains are denoted by empty black circles. }
\label{fig6}
\end{center}
\end{figure}

\begin{figure}
\begin{center}
\includegraphics[clip=on,width=80mm,angle=0]{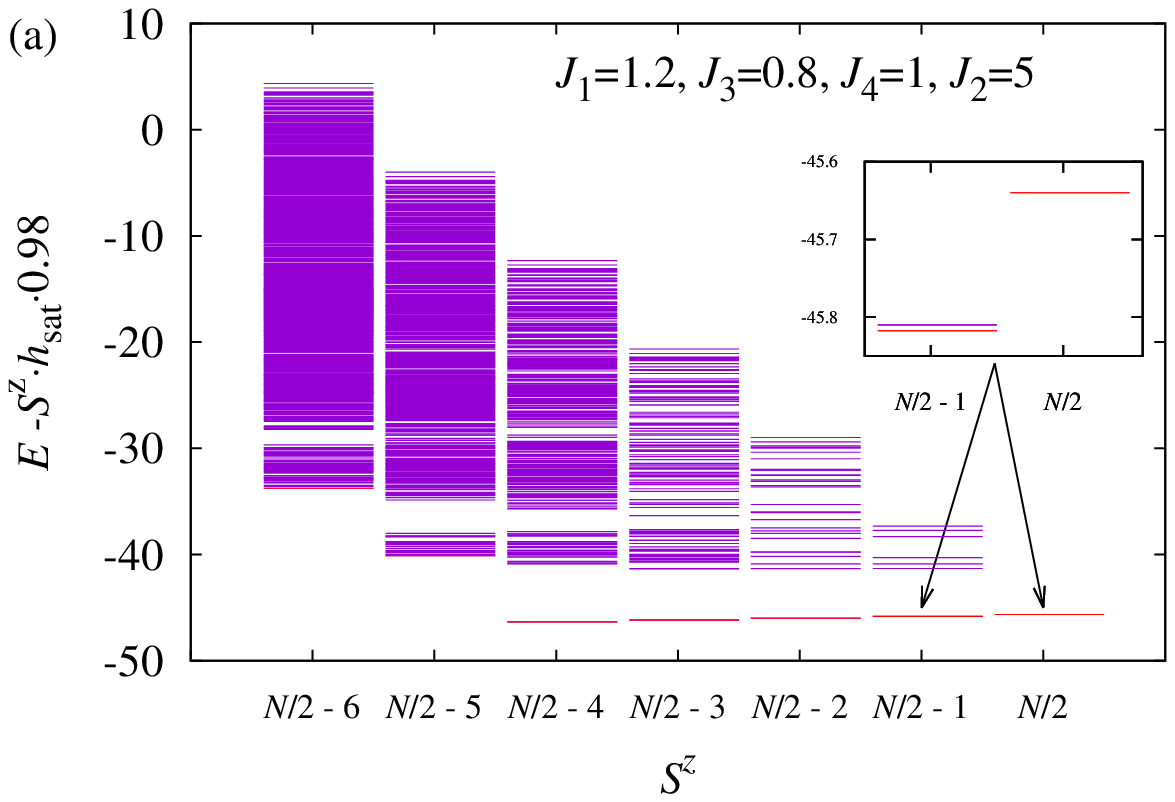}\\
\includegraphics[clip=on,width=80mm,angle=0]{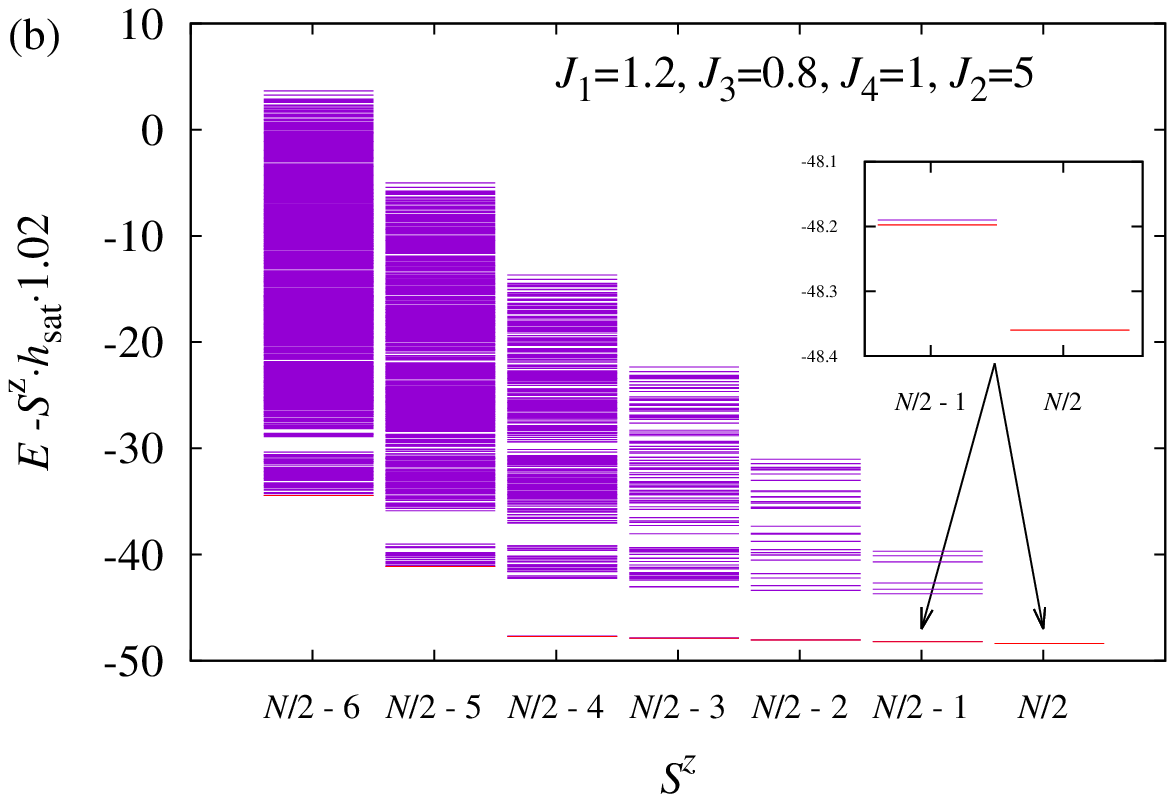}
\caption{{ (Color online) Energy levels of the distorted double-tetrahedral chain ($J_1=1.2$, $J_2=5$, $J_3=0.8$, $J_4=1$) in the subspaces with different value of $S^z$ operator in a magnetic field slightly bellow the saturation value (a) and in a magnetic field just above the saturation value (b).}}
\label{fig7}
\end{center}
\end{figure}

In Figs.~\ref{fig4}(a), ~\ref{fig5}(a) and \ref{fig6}
we also show the temperature dependences of the specific heat at high magnetic fields for ideal (red symbols) and slightly distorted geometries (green and blue symbols). { Temperature dependences of the entropy of ideal and distorted double-tetrahedral chain in strong magnetic field are depicted in lower panels of Figs.~\ref{fig4}, ~\ref{fig5}.}  Results for the corresponding effective models, see Eqs. (\ref{011}) and (\ref{015}), are denoted by thin black curves. The results of Jordan--Wigner fermionization for the effective model for the octahedral chain are denoted by black empty circles in Figs. \ref{fig3} and \ref{fig6}.

Now, we can discuss generic features, which arise as the consequence of small deviation from the perfect geometry of the lattice. First of all, deviation from the ideal geometry does not affect the noticeable change in width of the plateau
preceding the jump of magnetization to saturation (see Fig.~\ref{fig2} and Fig.~\ref{fig3}), but has an essential influence on the magnetization jump present for the ideal geometry. As one can see from right insets in of Fig.~\ref{fig2} and Fig.~\ref{fig3}, deviation from the ideal geometry {  together with finite temperature} leads to the smearing of the jump to saturation in the magnetization curve, which is another prominent feature of localized magnons both for the double-tetrahedral and the octahedral chains. For deformed initial chains the agreement between the effective theory given in  Eqs. (\ref{011}) and (\ref{015}) and the exact-diagonalization data becomes worse. The constructed effective theory within strong-coupling approximation reproduces the low-temperature behavior of magnetization in the vicinity of the saturation field only qualitatively and underestimates the width of the region where the low-temperature magnetization shows steep increase between two plateau values. 

As one can see from {  the upper panels of} Figs.~\ref{fig4},~\ref{fig5} and Fig.~\ref{fig6}, specific heat demonstrates a two-peak structure, which is typical for ideal geometry both for the double-tetrahedral and the octahedral chains. Low-temperature peak of the specific heat corresponds to the energy scale set by the degenerated manifold of states being ground states at magnetic fields near the saturation. Moreover, such characteristic extra low-temperature peak in the specific heat survives in distorted systems. In the case of the double-tetrahedral chain deviation from ideal geometry leads to the lifting of the degeneracy due to the chirality degrees of freedom and one of two-fold degenerated flat-band in energetic spectrum acquires a small dispersion. As a result, temperature dependence of the specific heat of deformed double-tetrahedral chain  in a magnetic field slightly below saturation value demonstrates three-peak structure. { It should be noted, that such three-peak structure of the specific heat appears exclusively at high magnetic fields just below the saturation value. The confirmation of this statement can be understood from the energy levels of the deformed double-tetrahedral chain in subspaces with different values of the operator $S^z$, $S^z=\frac{N}{2}, \frac{N}{2}-1, \ldots$, which are shown in Fig. {\ref{fig7}}. In a magnetic field a bit under the saturation value in subspace, for example, with one flipped spin($S^z=\frac{N}{2}-1$) system has degenerate low-lying excitations from the ground state to first-excited states. Such excited states are ground states in magnetic field below the saturation (see Fig. {\ref{fig7}}(a)), but when magnetic field is above the saturation value (see Fig. {\ref{fig7}}(b)) those states are not ground states anymore and therefore  do not contribute to the specific heat at very low temperature. On the other hand, the  explanation of appearance of an additional low-temperature peak in the specific heat can be understood from the temperature dependence of the entropy of double-tetrahedral chain in a strong magnetic field, see lower panels of Figs.~\ref{fig4}, ~\ref{fig5}. In the case of the ideal geometry of the double-tetrahedral chain ($J_1=J_3=J_4=1$, $J_2=5$) the high degeneracy of the localized eigenstates leads
to a residual ground-state entropy in magnetic field a bit bellow the saturation $S(h=0.98h_{sat})=\ln(2)$, which stems from the chiral degree of freedom. The deviation from ideal geometry of the lattice eliminates the degeneracy of the ground states and the temperature dependence of the entropy acquires a three peak shape.} An effective Hamiltonians are capable to reproduce not only the low-temperature peak of specific heat (ideal and deformed octahedral chains and ideal double-tetrahedral chain), but also the second peak in the specific heat of the deformed double-tetrahedral chain, see upper panels of Figs. \ref{fig4}, \ref{fig5} and Fig. {\ref{fig6}}. However, the constructed effective theory concerns to the low-temperature physics, and therefore does not reproduce the high-temperature maximum of the specific heat for both Heisenberg chains, considered in this paper.

\section{Conclusions}

\label{sec4}

\setcounter{equation}{0}

In conclusion,
we have examined the high-field low-temperatures properties of the spin-$1/2$ Heisenberg model on the double-tetrahedral and the octahedral chains both for ideal or distorted geometry of the lattice. Using the concept of localized magnons and strong-coupling approximation we proposed an effective theory to explain low-temperature thermodynamics of the considered systems in high magnetic field. It is shown, that the double-tetrahedral chain at high magnetic fields is described by unfrustrated spin-1 chain with nearest-neighbor interactions in a magnetic field while for the octahedral chain effective description is presented by the exactly solvable spin-$1/2$ isotropic $XY$ chain in a magnetic field. Moreover, the obtained effective Hamiltonians are much simpler than the initial ones: they are free of frustration, have smaller number of sites and refers to the reduced Hilbert space. New features which appears due to small deviation from ideal geometry at strong magnetic fields are i) the smearing of the perfect jump to the saturation in magnetization curve for both deformed chains and ii) appearance of the additional peak in the specific heat of the distorted double-tetrahedral chain in magnetic field just below the saturation value.

The developed approach can be also adapted to other frustrated quantum Heisenberg antiferromagnets with spin higher than $1/2$ and extended to the case of low magnetic fields. It is necessary to choose the dominant states with the least energy in the vicinity of zero magnetic and repeat all required calculations to construct an effective Hamiltonian. However, the aim of the present investigation concerns high-field regime, where manifestation of almost localized magnons is the most striking. 

{ It is worthwhile noting that Cu$_3$Mo$_2$O$_9$ \cite{Has08,Kur10,Mat12} can be presented as a double-tetrahedral chain, however with a different deformation type as was considered here. The construction of an effective theory in case of the deformation of equilateral triangles will be a task for the future study.}

\section*{Acknowledgments}

The author would like to thank O. Derzhko, T. Krokhmalskii and T. Verkholyak for invaluable discussions and comments. The present study was partially supported by the National Academy of Sciences of Ukraine under Grant for the Research groups of young scientists (Contract No. 7/2019) and by Grant of the President of Ukraine (Contract No. F82/221-2019).

\end{document}